\documentclass[12pt]{article}

\usepackage{scicite}
\usepackage{amsmath}
\usepackage{times}
\usepackage{graphicx}

\newenvironment{sciabstract}{%
\begin{quote} \bf}
{\end{quote}}

\title{Time-Resolved Open-Circuit Conductive Atomic Force Microscopy for Quantitative Analysis of Nanowire Piezoelectricity and Triboelectricity}

\author
{Yonatan Calahorra,$^{1\ast}$ Wonjong Kim,$^{2}$ Jelena Vukajlovic-Plestina,$^{2}$\\ Anna Fontcuberta i Morral,$^{23}$ Sohini Kar-Narayan$^{1}$\\
\\
\normalsize{$^{1}$Department of Materials Science and Metallurgy}\\\normalsize{ University of Cambridge, CB3 0FS, Cambrdige, UK}\\
\normalsize{$^{2}$Laboratory of Semiconductor Materials, Institute of Materials, School of Engineering}\\\normalsize{ Ecole polytechnique f\'ed\'erale de Lausanne (EPFL), 1015 Lausanne, Switzerland}\\
\normalsize{$^{3}$Institute of Physics, School of Basic Sciences}\\\normalsize{ Ecole polytechnique f\'ed\'erale de Lausanne (EPFL), 1015 Lausanne, Switzerland}\\
\\
\normalsize{$^\ast$To whom correspondence should be addressed; E-mail:  yc402@cam.ac.uk}
}

\date{}

\begin{document} 


\baselineskip24pt


\maketitle


\begin{sciabstract}
  Piezoelectric nanowires are promising materials for sensing, actuation and energy harvesting, due to their enhanced properties at the nanoscale. However, quantitative characterization of piezoelectricity in nanomaterials is challenging due to practical limitations and the onset of additional electromechanical phenomena, such as the triboelectric and piezotronic effects. Here, we present an open-circuit conductive atomic force microscopy (cAFM) methodology for quantitative extraction of the direct axial piezoelectric coefficients of nanowires. We show, both theoretically and experimentally, that the standard short-circuit cAFM mode is inadequate for piezoelectric characterization of nanowires, and that such measurements are governed by competing mechanisms. We introduce an alternative open-circuit configuration, and employ time-resolved electromechanical measurements, to distinguish between electrical generation mechanisms and extract the piezoelectric coefficients. This method was applied to nanowires of GaAs, an important semiconductor, with relatively low piezoelectric coefficients. The results obtained for GaAs piezoelectric coefficient, $\sim$0.4-1 pm/V, are in good agreement with existing knowledge and theory. Our method represents a significant advance in understanding the coexistence of different electromechanical effects, and in quantitative piezoelectric nanoscale characterization. The easy implementation will enable better understanding of electromechanics at the nanoscale.
\end{sciabstract}


\section*{Introduction}
The several decades long interest in semiconductor nanowires\cite{Guniat2019} (NWs) has brought focus to a topical niche - that of piezoelectric semiconductor NWs, pioneered by Wang \textit{et al.} \cite{Zhao2004piezoelectric,Wang2006Science,Wang2006Nano}, with potential applications in sensing, energy harvesting and logic\cite{Wang2010piezo,Wang2013triboelectric,Briscoe2015piezoelectric}. Three distinct electromechanical effects are manifested strongly in semiconductor nanowires: the high aspect ratio allows large elastic deformations, therefore enhancing the piezoelectric effect\cite{Zhao2004piezoelectric}, describing changes in surface polarisation due to applied strain; while the increased surface-to-volume ratio enhances interfacial effects such as triboelectricity\cite{Zhu2013toward,Choi2017}, relating to surface charge transfer upon contact with a dissimilar material; in addition, the combination of semiconducting and piezoelectric properties results in a unique electromechanical phenomenon known as the piezotronic effect\cite{Wang2010piezo,Fromling2018piezotronic}, whereby the height of a semiconductor energy barrier for charge carrier transport is changed due to mechanical pressure.\\ \indent
When considering electromechanical current/voltage generation from semiconductor NWs, both single NWs and NW ensembles or arrays have been considered\cite{Wang2014effects,Gogneau2014,Gogneau2016,Liu2016nanogenerators}. In particular, an abundance of conductive atomic force microscopy (cAFM) measurements have been reported on piezoelectric semiconductor NWs (mostly ZnO and III-N). In cAFM, the tip is scanned along the surface, while the current is recorded simultaneously, usually under applied bias. This mode of operation is also useful for other physical mechanism characterization, such as photovoltaics\cite{Mikulik2017conductive}. However, straightforward application of cAFM for electromechanical characterization of piezoelectric NWs is challenging. The two common variants for piezoelectric NW characterization are the short-circuit cAFM mode: recording the current generated by deforming a NW with an AFM tip  (see Fig. \nolinebreak\ \ref{fig:SEM_Scheme})\cite{Su2007generation,Wang2014effects,Liu2016nanogenerators,Alekseev2018piezoelectric}; and the loaded configuration (resiscope mode): recording the voltage developed across a resistor in parallel to the deformed NW\cite{Wang2006Science,Wang2010electricity,Gogneau2014,Gogneau2016,Jegenyes2018high}.\\ \indent
Although this is a widely studied topic and a commonly conducted experiment, very little attention has been given to the complete set of mechanically induced current flow mechanisms, namely piezoelectric, triboelectric and piezotronic. The combination of these effects is to be expected due to the nature of measurement involving dynamic forces and contact characteristics experienced by the NW-tip system, as the tip is scanned in and out of contact with NWs. Indeed, we are aware of only two reports attempting a quantitative analysis of the measured voltage or current \cite{Alekseev2018piezoelectric,Wang2010electricity}, to extract or compare the piezoelectric coefficients with theory - albeit with limited success.\\ \indent
The contrast between abundant experimental demonstrations of electrical generation from piezoelectric NWs and lack of quantitative discussion, is closely related to an uncertainty in the physical origins of the measured signals. In particular, the initial report on cAFM based generation from ZnO NW\cite{Wang2006Science}, was followed by alternative analysis of the results \cite{Alexe2008energy,Schubert2008finite}, and a rebuttal from Wang\cite{Wang2009energy}. Furthermore, flat ferroelectric samples have been studied by similar approaches\cite{Hong2014charge,Gomez2017piezo,Kwon2018direct}. In particular, a direct piezoresponse force microscopy (PFM) method, introduced by Gomez \textit{et al.}\cite{Gomez2017piezo}, was used to quantitatively extract direct piezoelectric coefficients; however, the nature of that method precludes its application to  nanomaterials, or materials of lower piezoelectric coefficients, as discussed below.\\ \indent
Herein, we report a new methodology to perform this experiment, enabling reliable extraction of the piezoelectric coefficients using cAFM. The prevailing method for quantitative analysis of piezoelectricity is PFM, which is generally not suitable for non-planar samples\cite{Calahorra2017mapping}, and characterises converse piezoelectricity (electrical to mechanical). We performed a detailed set of cAFM experiments at different conditions to isolate and distinguish the various current mechanisms in vertical pressure based cAFM measurements. Using scanning and ramp-mode cAFM in conjunction, we show that all three current mechanisms highlighted above are present in the case of GaAs NWs. We also show that the piezotronic and triboelectric currents could dominate the measurements to the point of obscuring piezoelectric data. We distinguish between short- and open-circuit configurations, and their practical applications in relation to cAFM based measurements. We suggest an open-circuit configuration and a time-resolved measurement methodology to extract the piezoelectric coefficients demonstrating good agreement to known values of GaAs. Obtaining these numbers for GaAs, a weak piezoelectric material, indicates promise to use with other piezoelectric materials. Our method offers a route for quantitative analysis of the direct piezoelectric effect. It is easily implemented, and can be extended to other piezoelectric nanostructures, as well as to characterize triboelectricity.\\ \indent
\paragraph*{Piezoelectricity, triboelectricity and piezotronics}
Research into nanoscale electromechanical phenomena has evolved from mere piezoelectricity to include a wide body of work related to the unique combination of piezoelectricity and semiconducting properties, coined as the piezotronic effect \cite{Wang2010piezo,Wang2014effects}, as well as to triboelectricity or contact electricfication\cite{Wang2013triboelectric}. The work presented here aims to distinguish these effects in cAFM.\\
\paragraph{Piezoelectricity}
Piezoelectricity, first described by Jacques and Pierre Curie, is the linear interrelation between electrical polarization and applied stress in non-centrosymmetric materials \cite{Curie1882}. There are four types of piezoelectric coefficients, correlating the electric displacement \textit{D}, electric field \textit{E}, stress field \textit{{T}} and strain field \textit{{S}}\cite{Damjanovic1998ferroelectric}    
\begin{align}
d_{ij}&=\left(\frac{\partial D_i}{\partial T_j}\right)^E=\left(\frac{\partial S_j}{\partial E_i}\right)^T\\
e_{ij}&=\left(\frac{\partial D_i}{\partial S_j}\right)^E=-\left(\frac{\partial T_j}{\partial E_i}\right)^S\\
g_{ij}&=-\left(\frac{\partial E_i}{\partial T_j}\right)^D=\left(\frac{\partial S_j}{\partial D_i}\right)^T\\
h_{ij}&=-\left(\frac{\partial E_i}{\partial S_j}\right)^D=-\left(\frac{\partial T_j}{\partial D_i}\right)^S
\end{align}
indices \textit{i,j} follow Voigt notation, \textit{e.g.}\nolinebreak\ $d_{ij}$ describes the $i$ component of polarization in response to stress applied in the $j$ direction, or alternatively strain in the $j$ direction in response to an electric field along the $i$ direction. The two alternative descriptions describe the direct (left-hand side equality) and converse (right-hand side) piezoelectric effect. The superscripts indicate differentiation under constant or zero fields. This has a bearing on the application of the relation to experimental conditions, depending on the boundary conditions. For example, a clamped piezoelectric structure operates under zero strain condition and will develop stress following the application of an electric signal. Alternatively, a structure clamped on one side operates under zero or constant stress. This is the operational condition of PFM, where an electric field is applied to the sample, and the strain is monitored - a direct measurement of the out of plane converse piezoelectric strain coefficient $d_{jj}$ (alternatively named the piezoelectric charge coefficient when considering the direct effect). The application of these equations to the direct piezoelectric effect will be discussed below in detail.\\ \indent
The coefficients are superpositioned with the linear mechanical or electrical equations to yield the complete linear electromechanical relations, for example  
\begin{align}
D&=\epsilon_T E + dT\label{eq:piezoelectricity1}\\
S&= d^tE + s_ET
\label{eq:piezoelectricity2}
\end{align}
where $\epsilon$ is the dielectric permittivity of the material, subscript $T$ represents constant stress, $d$ is the piezoelectric charge tensor, and $d^t$ is the transpose of $d$ (the interchangeably used piezoelectric strain tensor). $s_E$ is the elastic compliance at constant electric field. Notably, the piezoelectric charge coefficient, $d$, is used in short-circuit configuration (constant or zero electric field), where the piezoelectric element is treated as a current source. In an open-circuit configuration the piezoelectric voltage coefficient, $g$, is used - where the piezoelectric element is treated as a voltage source.\\ \indent
Piezoelectricity has many practical applications, including in sensing and actuation, however, there is a common thread to them: piezoelectricity is a decaying phenomenon. As such, only changes in stress yield useful electrical signals, and the main avenue of applications is in the AC regime.\\ \indent
\paragraph*{Triboelectricity}
Triboelectricity or contact electrification is a result of charge transfer between materials with different work-functions, where charge is transferred to equate the electrochemical potential upon contact\cite{Wang2013triboelectric}. Triboelectricity is not limited to semiconductors or metals, or crystalline materials, and is a property of soft and biomaterials as well\cite{Jing2018nanostructured}. The interest in nanoscale triboelectricity has peaked in the past decade due to the enhanced surface area of nanomaterials, significantly increasing the efficiency of triboelectric power generation\cite{Wang2013triboelectric,Jing2018nanostructured,Choi2017}. Similar to piezoelectricity, triboelectricity is also instantaneous. It occurs upon contact between the two materials, and decays with the interfacial electrochemical potential reaching equilibrium.\\
\paragraph*{Piezotronics}
The research into semiconductor nanowire piezoelectricity has brought into focus the combination of semiconducting and piezoelectric properties in a single material - the piezotronic effect.\cite{Zhou2008flexible,Wang2010piezo,Fromling2018piezotronic} Briefly, the main characteristic of a semiconductor is tunable conductivity through application of an electric field, as well as through physical contact with a different material. For example, in a pn junction or a metal-semiconductor contact, a depletion region in the semiconductor is formed in proximity to the interface. When it comes to piezoelectricity, depletion, or lower conductivity, is associated with higher piezoelectric responses. The piezotronic effect therefore describes the effect strain has on a semiconductor device, in particular the change of an energy barrier height due to changes in the interface charge, induced by the application of stress.\cite{keil2017piezotronic} The adequate physical treatment of this phenomena is not significantly different than that of non-ideal metal-semiconductor contacts having interface states within the band-gap, as suggested by Bardeen\cite{Bardeen1947surface,Sze2006}. The piezotronic effect takes the form 
\begin{equation} \label{eq:piezotronic}
    \Delta\phi_B=\phi_{B,PT}-\phi_{B,0}\sim\ P_{piezo}
\end{equation}
where the difference between the unperturbed barrier height ($\phi_{B,0}$) and the strained barrier height ($\phi_{B,PT}$) is proportional to the polarization ($P_{piezo}$).\\ \indent
There are two important distinctions between the piezotronic effect and the piezoelectric effect, from which it originates: i) the piezotronic effect does not decay, and the barrier height change (Eq. \ref{eq:piezotronic}) holds as long as the strain is held; ii) the piezotronic effect results in an exponential relation between the device current and the applied strain/stress, unlike piezoelectricity which is a linear effect by nature. This is due to the exponential relation between the current and the barrier height. This fact presents tremendous promise for piezotronic pressure and strain sensors, as we have recently demonstrated using GaAs NWs similar to the ones examined in this work\cite{Calahorra2019Highly}.
\paragraph*{Electromechanical cAFM measurements of semiconductor NWs}
The prevalent cAFM measurement of vertically aligned (as-grown) NWs is bending-induced current generation. This is due to two reasons, related to the geometry of the AFM tip/NW system: i) ideally, contact mode piezoresponse force microscopy (PFM\cite{Gruverman2006,Collins2019}) would have been desirable to measure the piezoelectric properties of the NW, and specifically the axial piezoelectric coefficient. However, NWs are brittle when considering contact mode operation, and therefore such reports are scarce, and require mechanical constraints\cite{Minary2012}. We have previously addressed this issue by developing a non-destructive PFM operation mode\cite{Calahorra2017mapping} and applying it to horizontal and vertical III-V NWs\cite{Calahorra2017exploring}. This limitation, together with the need to characterise the direct piezoelectric effect (rather than the converse effect in the case of PFM), has directed work towards non-contact mode electromechanical cAFM; ii) while slender NWs are easily bent, their axial stiffness is considerably larger. The bending (cantilever) and compression (axial) force constants of a typical NW are given by
\begin{eqnarray}
k_{bending} = \frac{3YI}{L^3} \longrightarrow \frac{Ya^4}{4L^3} \\
k_{axial} =  \frac{YA}{L} \longrightarrow \frac{Ya^2}{L}
\end{eqnarray}
where $Y$ is Young's modulus, $I$ is the second moment of inertia, $L$ the NW length, $A$ is NW cross section area, and $a$ is the edge of a hypothetical square cross-sectioned NW. For a moderate aspect ratio square cross-section NW, 1 $\mu$m long and 100 nm wide, with a Young's modulus of 100 GPa, we get $k_{axial} = 1000$ N/m, while $k_{bending} = 2.5$ N/m. It is therefore clear that bending NWs is generally much easier than compressing them, and hence the common cAFM NW characterization methodology measures current during bending while the tip is scanned across the NWs.\\ \indent
The common description for carrying out NW deformation-induced current generation measurements is as follows\cite{Wang2006Science,Wang2009energy}: as the tip is scanned across the sample and a NW is bent, the compressed and stretched sides of the NW develop a potential difference. A current flows when the tip is in contact with one of these sides. Interestingly, the common observation is that only one of the voltage polarities is associated with measurable current flow. This is attributed to the rectifying electrical properties of the tip-NW contact\cite{Wang2006Science}. Notably, this is in contrast to the common observation in NW ensemble device measurements, where two current/voltage peaks of opposite sign are associated with straining and relaxing the device\cite{Hasan2015hierarchical,Johar2018stable}. Moreover, the role of the rectifying contact is deemed even more fundamental. It was found that the existence of a rectifying contact is critical for electrical generation by the strain/compression of the piezoelectric semiconductor nanowire, to prevent a non-efficient current route (\textit{i.e.}\nolinebreak\ involving power dissipation within the NW) developing between the polarised regions of the NW\cite{Wang2006Science,Yang2009power,Gogneau2014,Gogneau2016,Liu2016nanogenerators}. The underlying reasoning could be that NWs with good electrical contacts are in fact highly doped, and their piezoelectric properties deteriorate. These points have been a part of the discussion regarding the origins of the measured signal\cite{Alexe2008energy,Schubert2008finite}.\\ \indent
\section*{Results and discussion}
\begin{figure}[t]
 \centering
 \includegraphics[scale=1.5]{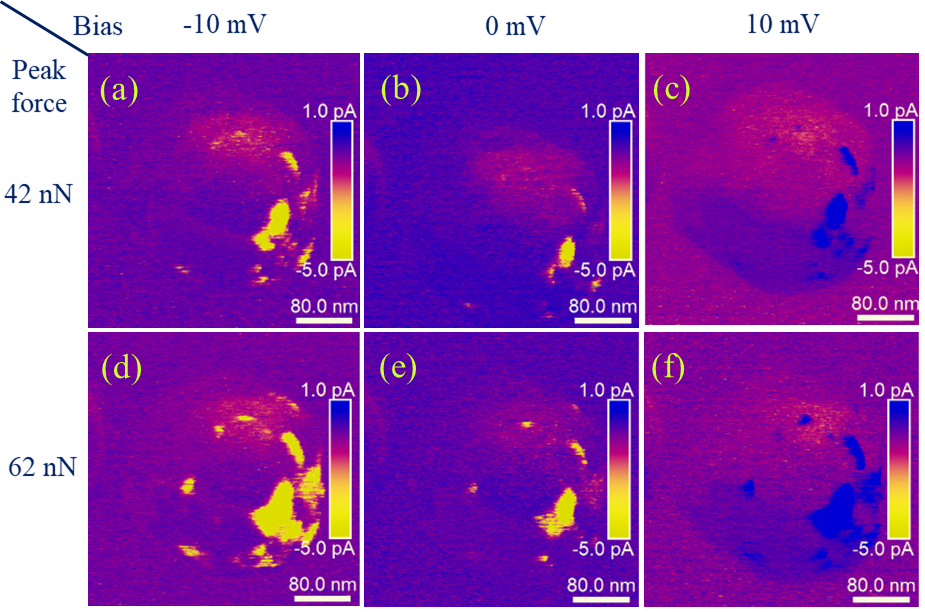}\caption{Bias influence on measured current map. Contact current maps obtained with 42 or (d-f) 62 nN peak-force, and sample-tip bias of -10/0/10 mV (a,d/b,e/c,f). The bias had a significant effect on measured current.}
 \label{fig:biassens}
\end{figure}

\begin{figure}[t]
 \centering
 \includegraphics[scale=1.5]{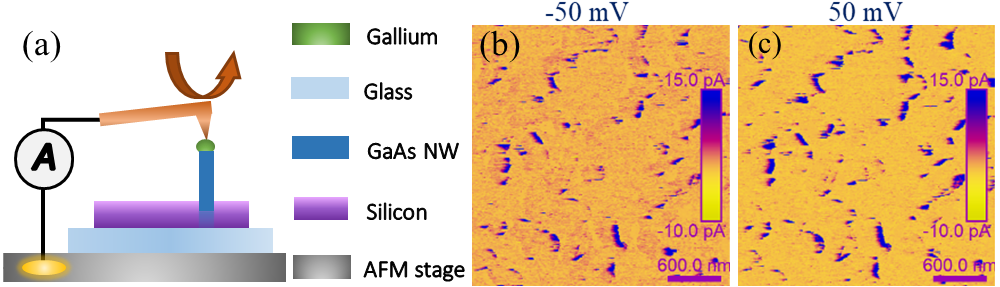}\caption{a) Open-circuit cAFM measurement schematic; b,c) consecutive current maps obtained from the sample while applying -50 [b], and 50 mV [c]. These images encompass both parasitic growth and nanowires which contribute to the signal. The isolation rendered the measurement not sensitive to the bias.}
 \label{fig:isolation}
\end{figure}

\paragraph*{Pressure-dependent contact current }
\begin{figure}[t]
 \centering
 \includegraphics[scale=1.6]{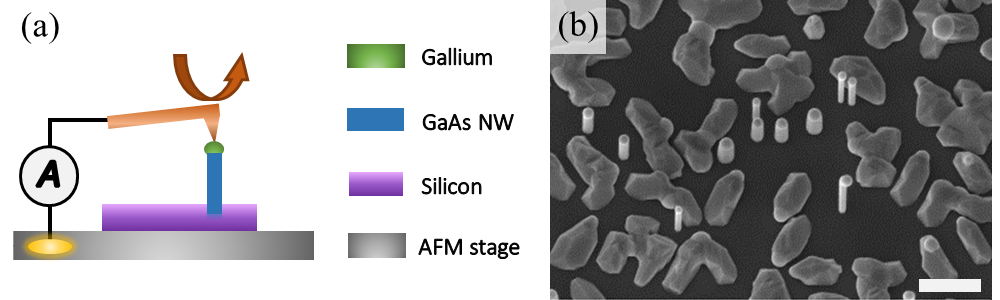}
 \caption{(a) Measurement schematic, showing the AFM cantilever descending atop the NW, and the current measured; (b) The sample used in these measurements viewed by SEM in an angle of 20$^{\circ}$. Scale bar is 1 $\mu$m.}
 \label{fig:SEM_Scheme}
\end{figure}
Figure  \ref{fig:SEM_Scheme}a shows a schematic of the experimental procedure, where the AFM tip is scanned in PeakForce mode\cite{QNM} along the surface, while recording the resulting current, referred to as the PF-TUNA mode. Figure \ref{fig:SEM_Scheme}b shows a scanning electron microscope  (SEM) image of the examined NW sample. Nanowire growth time was controlled to maintain a relatively short length, rendering the NWs more stable under AFM scanning. Noticeably, alongside the NWs, parasitic growth dominates the silicon surface, on which the NWs were grown by molecular beam epitaxy (MBE).\\
\begin{figure}[t]
 \centering
 \includegraphics[scale=1.6]{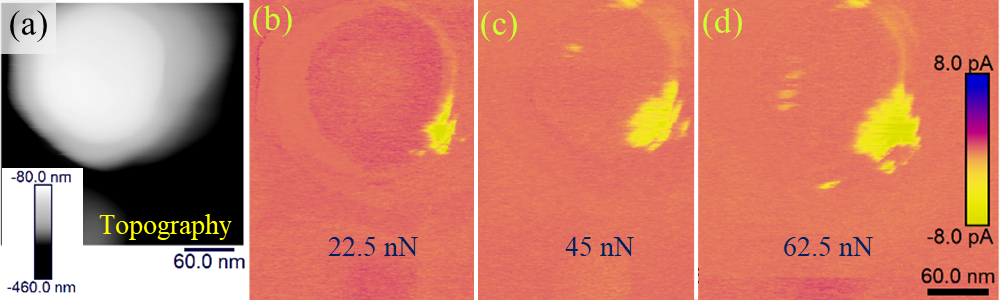}\caption{Scanning peak-force influence on measured short-circuit current map (a) Single NW height channel; (b,c,d) contact current maps obtained with 22.5, 45, 62.5 nN peak-force. The active area increased with applied force.}
 \label{fig:pressuresens}
\end{figure}
\indent Figure \ref{fig:pressuresens} shows a set of PF-TUNA current maps obtained from a NW (topography in black and white), with increasing peak-force (the maximum mechanical load, see Methods Section) values of 20-60 nN. The electrical response area increased with the mechanical stimulus, however the highest current was measured at 45 nN peak force. This indicates that the relation between the measured current and applied force is not straightforward. Furthermore, the current was not correlated to the top of the NW, but to a region along its side, found to be consistent throughout the measurements. This result could be explained with an increased deformation of this side, however the measured current was found to be uncorrelated with the deformation maps {(Supporting Information Figure S1)}, and this result recurred throughout this work. Moreover, this observation was reinforced through measurements obtained from the sample in general: zero-bias current is observed throughout the sample, upon contact with the surface, and the parasitic GaAs growth (see Figure S2). Bearing in mind that the parasitic growth is not expected to produce a significant piezoelectric signal, and is irregularly deformed, we conclude that the current observed is not piezoelectric per se. The following set of experiments explore the two additional electromechanical current mechanisms.\\
\paragraph*{Bias-dependent contact current}
To further explore the nature of the measured current, a small bias was applied to the sample during scanning. Figure \ref{fig:biassens} shows current maps obtained with lower/higher peak force setting (top and bottom) and three biasing conditions: negative, zero and positive. The bias used was  $\pm10$ mV. The middle column of Fig.\nolinebreak\ \ref{fig:biassens}, is similar to Fig.\nolinebreak\ \ref{fig:pressuresens}, where increased peak force results in broader current ``hot-spots". However, the bias sign was found to dictate the measured current sign. This is another indication that the measured hot-spot current is not piezoelectric. The increase in current hot-spot area with pressure could indicate that the dominant mechanism is improvement of the contact, and not piezoelectricity. We note that, following this experiment, we have found that there is an electrical ground imbalance in the instrument of about 3.5 mV, which probably contributes to the current measured in zero bias. \textit{I-V} measurements taken from the current hot-spots and the low-current areas on top of the NW show a complementary picture where the low-current areas show a rectified \textit{I-V} curve, while the hot-spots show better conductivity. This is most likely related to the geometry of the NW and tip, and a local barrier reduction due to Schottky effect {(Supporting Information Figure S3)}. In these experiments, there where no conclusive findings relating the current to applied pressure, \textit{i.e.} the piezotronic effect. \\
Furthermore, since GaAs is considered here, there might be an optoelectronic contribution to the current through carrier generation by the AFM laser, as demonstrated recently by Alekseev \textit{et al.}\cite{Alekseev2018piezoelectric}. It is therefore of interest to eliminate steady-state (ohmic/piezotronic) contributions to the current. 
\paragraph*{Open-circuit current measurements}
In order to reduce unwanted steady-state current contributions, we introduced a standard microscope glass slide between the sample and the AFM stage; this is schematically shown in Fig.\nolinebreak\ \ref{fig:isolation}a. Notably, the existence of a ``steady-state" electrical contact (\textit{i.e.\nolinebreak} ohmic or rectifying contact) is not necessary for the observation of generated current/voltage in piezoelectric generators\cite{Hasan2015hierarchical,Johar2018stable}. This is obvious when considering the issue outside the context of piezoelectric semiconductors: the best piezoelectric materials are insulating, and therefore do not have good electrical contacts. If so, the central role of the Schottky contacts in cAFM measurements of electromechanically induced current (as mentioned above) implies that this current is indeed not purely due to piezoelectricity. Therefore, we set out to examine cAFM-induced current in an open-circuit configuration.\\ \indent
Figure \ref{fig:isolation}b,c provides complementary results to Figure \ref{fig:biassens}, obtained under open-circuit conditions. The current was found to be independent of applied bias, in a striking contrast to the standard configuration. This implies that steady state currents were dominating the measurements before. Nonetheless, the measured current was still not negligible, and maintained the hot-spot characteristics. This result is intriguing: i) if ohmic/piezotronic contributions are eliminated, then triboelectric and piezoelectric  currents are possible; ii) considering the uncorrelated nature of the deformation maps and the current maps, the reasonable conclusion is that triboelectricity is the dominant effect in the generation of current hot spots in the open-circuit regime. This result is in good agreement with recent reports of the co-existence of triboelectricity and photovoltaic current generation in a III-V material probed by cAFM\cite{Sharov2019inp}. The question therefore remains: is it possible to record the piezoelectric generation arising from a single NW?\cite{Alexe2008energy,Schubert2008finite,Wang2009energy}
\paragraph*{Time-resolved current measurements}
\begin{figure}[thb]
 \centering
 \includegraphics[scale=1.6]{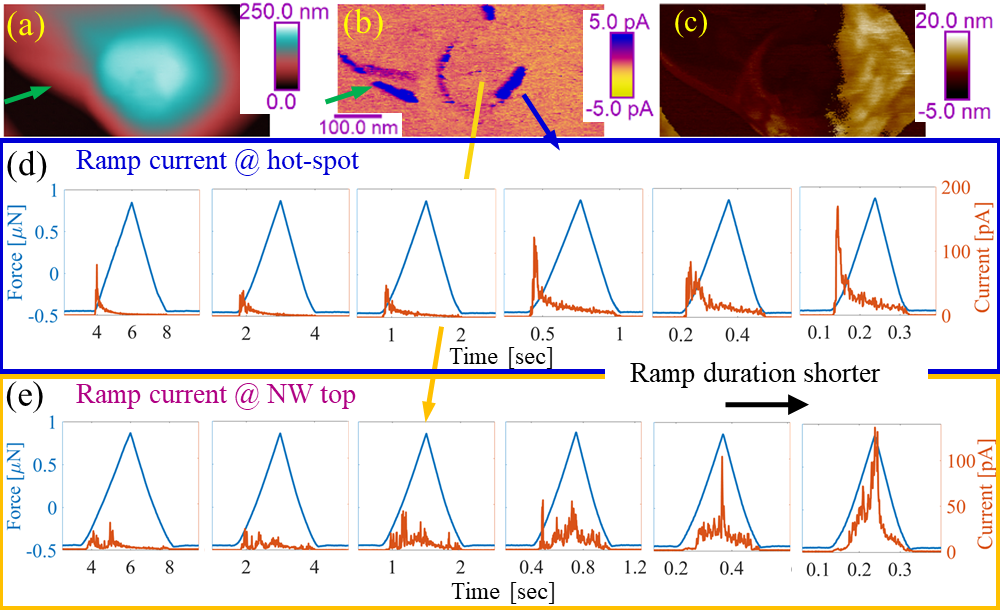}\caption{Mechanical ramping induced open-circuit measured current: time-resolved. (a) Topography (b) current map and (c) deformation map of a NW later used to obtain feedback-controlled time-resolved current from the high (d) and low (e) current locations; (d,e) a series of time-resolved current measurements obtained by mechanical ramping on-top the same position, to the same maximal force, with varying ramping rate. Note the current axis for (e) and (d) are not identical. low- and high-current positions yield distinct time-resolved current characteristics, interpreted as piezoelectric and triboelectric contributions.}
 \label{fig:rampcurrent}
\end{figure}
In order to further investigate electromechanical current generation, we moved on to explore the time-resolved characteristics obtained through mechanical ramping in open-circuit configuration. This has several advantages compared to the scanning mode: i) the currents in scanning mode are averaged over significant periods of time (\textit{e.g.}\nolinebreak\ the entire contact time or the entire peak-force period). Therefore time-resolved measurements provide additional information; ii) ramping is less destructive than scanning and larger forces can be used, allowing stronger indentations. Here we drove the cantilever into the sample up to a force of $_\sim$ 1300 nN, which is roughly 10 times larger than the peak-force used during scanning.\\  \indent
Figure \ref{fig:rampcurrent} shows time-resolved current measurements obtained from a current hot-spot and from the top of the NW, where generally low currents were found. The top part of the image shows the scanning results, demonstrating once more the lack of correlation between deformation and current. Low- and high-current spots were chosen, and a series of mechanical ramps with a preset maximal force were executed, while changing the ramping rate from 20 nm/sec (left-hand side) to 505 nm/sec (right-hand side). The tip was completely out of touch with the sample after every measurement, as evident from the force curve plateau.\\ \indent The two ramping points gave rise to distinct traits: the current hot-spots show a current peak measured directly upon tip-sample contact, and a subsequent decay of the current. Conversely, the low-current NW top, gave rise to a current peak which is correlated with the trend of the applied force - most evident in faster (shorter) ramps. These findings support the assumption that triboelectric current is the dominant mechanism explaining the current hot-spots, considering the instantaneous onset of current and the lack of correlation with deformation (green arrows in Fig.\nolinebreak\ \ref{fig:rampcurrent}). Possibly, the geometry of the tip and sample results in spatial locations which favour the onset of triboelectric current, for example where sharp features result in a locally enhanced electric field supporting charge transfer.\\\indent
The current measured on top of the NW, with good correlation to the applied force (Fig.\nolinebreak\ \ref{fig:rampcurrent}e), is an indication of piezoelectric generation. This follows the elimination of DC currents and triboelectric currents, as the dominant mechanisms in this position. However, there are still ambiguities in understanding this current: firstly, measurement of GaAs under the red AFM laser induces an inherent contribution from carrier generation,\cite{Alekseev2018piezoelectric} and it is possible that the contact current here is affected by that mechanism. Indeed, overlaying the currents in Fig.\nolinebreak\ \ref{fig:rampcurrent}d,e indicates that a possible background current is present during the measurement, apart from the peaks (see supporting Figure S4). Secondly, the lack of an opposite current peak is intriguing: piezoelectricity is linear, and therefore it might be expected that both positive and negative currents will appear upon compressing and releasing the NW.  
\paragraph*{Extraction of piezoelectric coefficients}
\begin{figure}[th]
 \centering
 \includegraphics[scale=1.4]{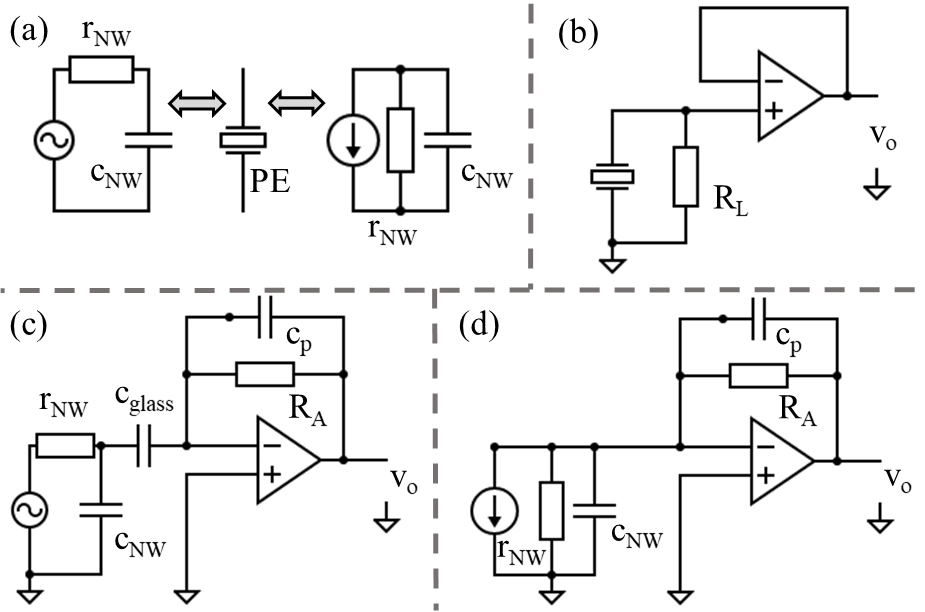}
 \caption{Electronic circuits depicting the different configurations for piezoelectric generation measurement a) the piezoelectric element as a current or voltage source, with the NW capacitance and resistance connected in the appropriate manner; b) the loaded resiscope mode, where neither open-circuit voltage nor short-circuit current are measured, with $R_L$ as the load and the op-amp as a voltage follower; c) the open-circuit configuration, realizing a voltage differentiator, with the glass slide capacitance $c_{glass}$, and the feedback resistor and capacitor, $R_A,\ c_p$; d) the short-circuit configuration.}
 \label{fig:circuits}
\end{figure}
For extracting the direct piezoelectric coefficient, mirroring (converse) PFM, it is necessary to quantify this measurement. We examine the different configurations in their electronic context in Figure \ref{fig:circuits}. The operation of a generator is influenced by its load, hence the piezoelectric element acts as a voltage source under open-circuit conditions, and as a current source under short-circuit conditions (Fig.\nolinebreak\ \ref{fig:circuits}a). Analysis of the loaded configuration is not straightforward, and indeed, the operation efficiency of piezoelectric energy harvesters (as any electrical generator) is affected by the load \cite{Briscoe2015piezoelectric,Crossley2015} (Fig.\nolinebreak\ \ref{fig:circuits}b). Inversely, the open-circuit and short-circuit configurations are straightforward for (theoretical) analysis, and correspond to the piezoelectric element acting as a voltage and current source correspondingly (Fig.\nolinebreak\ \ref{fig:circuits}c,d). In particular, our implementation of the open-circuit configuration, with a capacitor in series to the device, is a voltage differentiator, where the measured current maintains
\begin{equation}
    i_{measured}=c_{glass}\frac{dv}{dt}\label{Eq:differentiator}
\end{equation}
The other components shown in Fig.\nolinebreak\ \ref{fig:circuits} are the NW internal resistance ($r_{NW}$), which is expected to be high in our case, and capacitance ($c_{NW}$), which is determined by the surface area of a single NW and therefore very low ($\sim{}10^{-18}$ F). $R_A$ is the current to voltage converter feedback resistor, and $c_p$ is the feedback/stray capacitance. According to AFM manufacturer $R_A\simeq5$ G$\Omega$ and $c_p\sim{}$ fF - accounting for about 10 kHz operation bandwidth \footnote{private communication with Bruker}. The glass capacitor used here is comprised of an AFM sample holder plate (1.25 cm in diameter), and the AFM stage, with a 1 mm glass slide in between. This result is $c_{glass}=5.3$ pF, assuming relative permittivity of 5\cite{Glass}. Therefore $c_{glass}$ is significantly larger than the other capacitors in the circuit, and comes into play. Furthermore, the time-resolved experiment does not involve rapid changes to the equivalent circuits, since the tip is in consistent contact with the NW, therefore there are no significant changes expected in the contribution of the op-amp current bias to the signal\cite{Alexe2008energy}. 
\paragraph*{Short-circuit configuration}
Short-circuit across the NW (3-axis) indicates that (for simplicity, considering axial fields and displacements alone)
\begin{align}
E_3&=0\\
D_3&=P_3=d_{33}T_3
\end{align} 
since there are no fields outside the nanowire, the boundary conditions for the displacement field $D$ hold $D=\pm Q/A$, where $A$ is the effective electrode area, and $Q$ the free interfacial charge, neutralising the polarisation charge, brought by the deformation. Ideally, the current measured in the experiment is the movement of charges to neutralise the polarisation - a generated current (Fig.\nolinebreak\ \ref{fig:circuits}d).\\ \indent
The dimensions of the NW measured in Fig.\nolinebreak\ \ref{fig:rampcurrent} are $L_{NW}=210$ nm and $r_{NW} = 40$ nm. We have tried to take the significant uncertainty in the diameter of the NW (lateral AFM measurement), of about $\pm 20$ nm, into account. The piezoelectric charge coefficient  of GaAs is $d_{33}\simeq 1.5-2.5$ pm/V (pC/N)\cite{Calahorra2017exploring}. If so the application of 1300 nN should result in
\begin{equation}
    Q=d_{33}\cdot F_{app} = 2.5 \cdot 10^{-12} \cdot 1300 \cdot 10^{-9} = 3.25\cdot 10^{-18}\ \mathrm{C}
    \label{eq:polarisationcharge}
\end{equation}
Let us assume this force is reached within 0.1 sec, corresponding to the fastest ramps in Fig.\nolinebreak\ \ref{fig:rampcurrent}. In that case, the current measured will be $Q/\Delta t=3.25\cdot10^{-17}$ A. This current is in fact much lower than the noise level in our system ($\sim$100 fA). Moreover, even if we consider a material with $d_{33}= 100$ pC/N, and a force 100 times stronger, the current will still be roughly within the noise limit. Indeed, Gomez \textit{et al.}\nolinebreak\ circumvented this limitation by using an external current amplifier in their direct-PFM application\cite{Gomez2017piezo}. They measured relatively strong piezoelectric materials, having a flat topography.\\ \indent
We therefore conclude that the short circuit AFM configuration, where the piezoelectric charge coefficient dominates\cite{Calahorrabook}, is mostly not suitable for the measurement of piezoelectric current generation. When considering that it is in fact a single NW measurement, this result is not surprising, and corresponds with previous criticism of single NW current generation\cite{Alexe2008energy,Schubert2008finite}. This also corroborates to our observation and conclusion above that other current mechanisms prevail in this configuration, and are probably what has been measured in previous work utilizing short-circuit cAFM. The piezoelectric charge (strain) coefficient, $d$, is directly measurable under the application of constant electric field and monitoring of strain, \textit{i.e.}\nolinebreak\ PFM mode.\\ 
\paragraph*{Open-circuit configuration}
We move forward to the open-circuit configuration, where the boundary conditions hold $D_{boundary}=0$\cite{Li2002effect,Eliseev2015finite}. The relevant coefficient under zero displacement field is $g_{33}$, which is the ratio of electric field to applied stress - the piezoelectric voltage constant. As discussed above, in our configuration the voltage is not measured directly but the current in the circuit is recorded. This current is linearly related to the voltage time derivative (Eq.\nolinebreak\ \ref{Eq:differentiator}).\\ \indent
\begin{figure}[th]
 \centering
 \includegraphics[scale=1]{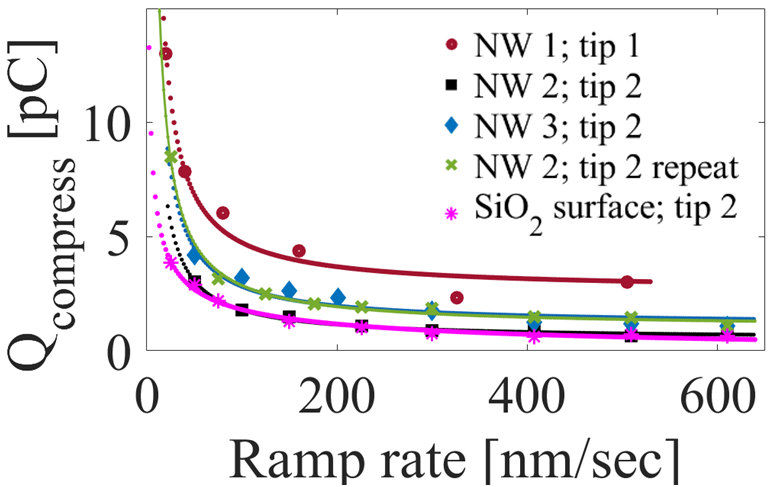}
 \caption{Integrated current during force application as a function of cantilever ramping rate (marks), and power law fitting of the data to extract the rate independent part. The results were obtained from several NWs, as well as a control experiment measured on the oxide covered surface. Doted lines are fittings according to Eq.\hspace{0pt} \ref{eq:powerlaw}. Tip 1 and Tip 2 are both \textit{Adama} AD-40-AS, with force constants of 38 and 27 N/m, correspondingly.}
 \label{fig:Qrise}
\end{figure}
To extract the voltage related to the application of the force we integrate the current. We focus on the charging part of the time-resolved force-current curves (ramp approach). Figure \ref{fig:Qrise} shows that charge for several NWs, as a function of cantilever ramping rate, obtained by numerically integrating the current up to the point where peak-force is reached (Fig.\nolinebreak\ \ref{fig:rampcurrent}). As discussed earlier, there still might be parasitic mechanisms contributing to the measurement. Assuming these are linear with the contact time, the ``piezoelectric" component should be mostly dependent upon the force, rather than the time. The integrated current is therefore fitted to
\begin{equation}
    Q=a/rate+q_{PE} \label{eq:powerlaw}
\end{equation} where the $1/rate$ term is proportional to contact duration, and therefore $q_{PE}$ is the time independent charge. The results for NW \#1 indicate that for the 1300 nN force, the charge accumulated on the capacitor was 2.68 pC. As mentioned earlier, $c_{glass}=5.3$ pf. If so, the voltage drop on the capacitor is ${q_{PE}/c_{glass}}$ which yields 0.505 V. Using the NW geometry mentioned above, we find the associated electric field is
\begin{equation}
    E_{piezo}=V_{piezo}/L_{NW}
\end{equation}
which yields $E_{1300 nN}=2.4$ V/m. The associated stress is given by $T=F/A$, yielding $T_{1300 nN} = 258$ N/m$^{2}$. Calculating the piezoelectric voltage coefficient, $g_{33}$ ($E/T$), yields $g_{33,GaAs}=0.0091$ Vm/N. The charge and voltage coefficients hold
\begin{equation}
    d=g\cdot\epsilon_0\epsilon_r
\end{equation}
using $\epsilon_{r,GaAs}=13$,\cite{GaAs} we get a value for the charge coefficient $d_{33,GaAs}=1.05$ pC/N. A similar analysis for two other NWs (NW 3 and the higher value for NW 2, see Fig.\nolinebreak\ \ref{fig:Qrise}) yields $d_{33,GaAs}=0.4-0.55$ pC/N. These values are in good agreement with known values and theoretical calculations for GaAs of 1.5-2.5 pC/N. We note that the experimental results are somewhat lower than theory. A possible reason is that the force could contribute to bending the NW and not only compressing it, effectively reducing the apparent mechanical to electrical coupling, resulting on a lower coefficient calculated. Furthermore, we see the uncertainty in diameter as the main contributor to the experimental error. The coefficients are proportional to the diameter square, and hence the associated error in the coefficient is double that of the diameter, which we asses as 20\%  - yielding $\pm\ 40\%$ uncertainty in the coefficients.\\ \indent
The time-resolved nature of this analysis determines the limits of the method, to be maintained for a valid analysis: 1) the ramping rate needs to be fast enough for resolving the contribution of the piezoelectric charging, and diminishing the contributions of other mechanisms (Fig.\nolinebreak\ \ref{fig:Qrise}); 2) the ramping rate needs to be slow enough for the system to respond to the changes. In our case with $C\simeq5$ pF, and the NW resistance assessed in M${\Omega}$-G${\Omega}$, we have a time constant of $\tau=RC\simeq10\ \mu$S - $10$ mS. This indicates that even our fastest ramps (100-200 mS long) are slow enough for the piezoelectric NW to charge the capacitor. This also indicates that the rapid peak-force operation ($\sim$1-2 kHz), might be too fast to measure the developed voltage, explaining the lack of measured signal in scanning mode.\\ \indent 
We performed several control experiments to validate the measurements (Supporting Information Figure S5): i) the open-circuit time-resolved procedure on the native oxide layer atop the Si growth substrate away from the NW growth area (Fig.\nolinebreak\ \ref{fig:Qrise}), as well as on an unrelated sample with a top ITO electrode; ii) short-circuit time-resolved measurements of the ITO electrode; iii) short-circuit time-resolved measurements of a different GaAs NW sample, where the Si substrate is undoped. It was found that the silicon oxide layer gave rise to an electromechanical signal, which was lower, though comparable, to the integrated current during NW measurements. Although native silicon oxide is not piezoelectric, it was recently demonstrated that the AFM electromechanical apparatus is sensitive to flexoelectricity - an electrical response brought by internal strain gradient and vice-versa. Abdollahi \textit{et al.}\nolinebreak\ have shown that AFM tips induce non-uniform electrical fields to layers during PFM operation, which results in the onset of the converse flexoelectric effect - manifested as an apparent piezoelectric signal\cite{Abdollahi2019converse}. Furthermore, flexoelectricity can contribute to tip-enhanced photovoltage generation\cite{Yang2018flexo}. It is very likely that in our case, where the direct piezoelectric effect is examined, flexoelectricity is responsible for the signals obtained from the oxide dielectric. The sharp tips used in this work (10 nm tip radius) corroborate with the quantitative analysis presented by Abdollahi and co-workers. Notably, piezoelectricity and flexoelectricity can co-exist in piezoelectric materials. In that sense, vertical NWs are advantageous compared to films, as the deformation of a rod would be more uniform than that of a layer, when subjected to a vertical load by an indenter of similar dimensions.\\ \indent
Measurements on top of an ITO electrode in the open-circuit configuration (Supporting Information Figure S5) gave rise to the decaying curves, similar to those measured around the current hot-spots, attributed to triboelectricity. In the short-circuit configuration, the results were entirely different: the current was 10-100 times larger, and in the opposite polarity. Finally, measurements of a similar GaAs NW grown on a non-conductive Si substrate gave a flat reading. These controls show that the current measured in the differentiator configuration is not an artefact of the system.\\ \indent
This method bears great implications for piezoelectric analysis at the nanoscale. As mentioned above, only a handful of reports attempt the extraction of piezoelectric coefficients from the measured data. On the one hand, using the short-circuit configuration, the generated currents are too low, and using the measurements generated by other mechanisms yield faulty results. On the other hand, the loaded configuration (resiscope mode), is neither open- nor short-circuit, making analysis more complex\cite{Wang2010electricity}. This leaves our open-circuit configuration as a promising alternative for the analysis of piezoelectricity in nanoscale structures, providing a applicable route to decouple various electromechanical effects.\\ \indent
\section*{Summary}
To conclude, we provide a comprehensive theoretical and experimental analysis for AFM-based piezoelectric generation. We have demonstrated for the first time that three electromechanical current mechanisms, piezotronic, piezoelectric and triboelectric, are present in the common short-circuit configuration, and that triboelectricity and piezotronic effects dominate the measured current. These findings help settle controversy related to this type of measurement\cite{Alexe2008energy,Schubert2008finite,Wang2009energy}, by showing that indeed piezoelectric current generation is unlikely the effect measured, while other electromechanical effects are present. We introduced an open-circuit configuration, circumventing fundamental limitations we found in the commonly used short-circuit configuration. Through time-resolved measurements of the generated current (effectively voltage differentiation), we extracted the piezoelectric voltage coefficient, which is related to the piezoelectric charge coefficient. This methodology and analysis procedure will allow characterization of nanoscale piezoelectric materials and advance the efforts of improving piezoelectric actuators, sensors and energy harvesters. It can be generalized for examining triboelectric, flexoelectric and piezotronic devices, depending on the dominant mechanism. 
\section*{Materials and Methods}
\paragraph*{Nanowire Growth and Morphology}
Ga-catalyzed GaAs NWs were grown on doped silicon by molecular beam epitaxy (MBE), in a process similar to previous reports\cite{Colombo2008ga,Matteini2015wetting,Matteini2016}. See Supporting Information Section S1 for full details.
\paragraph*{AFM characterization}
For AFM characterization, the sample back side was sputtered with gold, and mounted on a circular metallic AFM holder using silver paint. Atomic force microscopy was carried out using a \textit{Bruker} Dimension Icon microscope, using the PF-TUNA mode, through scanning, I-V and I-Z (ramping) operation. In PF-TUNA scanning mode, the tip is periodically oscillated above the surface such that it form intermittent contact with the sample, up to a specified ford, \textit{i.e.}\nolinebreak\ the peak-force.  The tips used were MESP-RC-V2 (coated silicon) by Bruker, and Adama AD-40-AS (diamond tips, to prevent tip damage; all time-resolved measurements reported were done using these tips). To realise an open-circuit configuration an optical microscope glass slide was introduced between the conductive AFM stage and the sample holder. The general experimental procedure was: i) locating a NW through a $\sim$10-15 $\mu$m scan, zooming-in, switching to ramp mode in the desired location, where the current was recorded during ramping with controlled rate and force.
\paragraph*{Analysis}
Growth analysis was done using SEM (Zeiss MERLIN). AFM results were analysed using NanoScope Analysis software, and MATLAB was used for calculations and presentation of results.
\bibliography{current.bib}

\bibliographystyle{Science.bst}

\section*{Acknowledgments}
Y.C.\nolinebreak\ and S.K-N.\nolinebreak\ are grateful for support from ERC Starting Grant (Grant No. ERC–2014–STG–639526, NANOGEN), as well as Henry Royce Institute - Cambridge Equipment grant EP/P024947/1 and the Centre of Advanced Materials for Integrated Energy Systems "CAM-IES" grant EP/P007767/1. W.K., J.V.P.\nolinebreak\  and A.F.i.M.\nolinebreak\ thank SNF for funding through the NCCR QSIT and H2020 via project Indeed and Nanoembrace.

\section*{Supplementary materials}
Supplementary Text\\
Figs. S1 to S5\\
\paragraph*{Section S1. Experimental: NW growth}
GaAs nanowires were obtained by the Ga-assisted growth method on silicon\cite{Colombo2008ga,Matteini2015wetting,Matteini2016}. For this, four inch $<$111$>$ p-doped Si wafers with a resistivity of $<$ 0.03 $ \Omega\cdot$cm (doping $10^{18}$ cm$^{-3})$ were diced into 35 $ \times $ \ 35 mm$^2$ square chips to fit in the MBE sample holder. The native oxide of the silicon was first removed by dipping the sample in a buffered HF solution (7:1) for 5 min. Subsequently, a 1.1 nm oxide was grown using 200 W, 200 sccm O$_2$ plasma power in TEPLATM GigaBatch for 30 seconds. The final oxide thickness was measured by spectroscopic ellipsometry. The prepared Si chips were introduced into the ultra-high vacuum (UHV) environment MBE machine (DCA P600) and subsequently annealed at 500 $ ^{\circ} $ C for 2 hours in UHV to ensure a pristine surface free of water and organic molecules. They were then transferred to the growth chamber where they were de-gassed at 770 $ ^{\circ} $ C for 30 min to further remove any possible surface contaminants. The sample was then brought to the growth temperatue, 634$^{\circ} $ C.  Then, Ga was pre-deposited for 10 min at beam equivalent pressure (BEP) of 1.4$ \times  10^{-7}$ Torr (corresponding to a nominal deposition rate of 1.1  $\mathring{A}$/s). After than, the As$_4$ source was opened for 20 min at BEP of 2$ \times  10^{-6}$ Torr. No dopants were introduced and hence the NWs were nominally undoped. Identical nanowires were used in the publication by Calahorra \textit{et al}.\cite{Calahorra2019Highly}.\\

\begin{figure}[ht]
\begin{center}
    \includegraphics[scale=1.8]{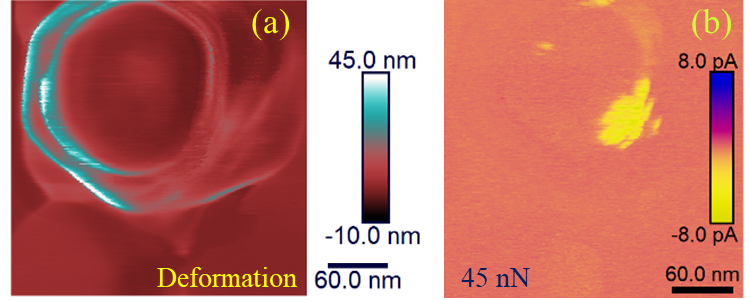}\\
\end{center}
 Fig. S1. (a) Deformation map and (b) Contact current map of the NW in Figure 2 in the main text. Demonstrating the lack of correlation between the high current areas and deformation. The deformation and current spatial characteristics remained similar throughout the measurements, even following sample rotation. They seem to be mostly linked to tip geometry.
\end{figure}

\begin{figure}[ht]
\begin{center}
    \includegraphics[scale=2]{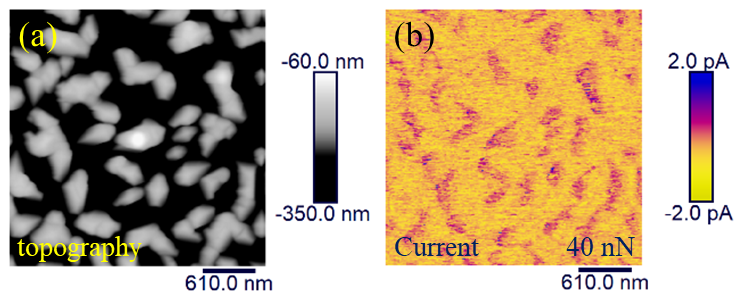}\\
\end{center}
Figure S2: (a) Topography and (b) Contact current map of typical large-area view of the sample. Demonstrating that current is not exclusive to NWs.
\end{figure}

\begin{figure}[ht]
\begin{center}
    \includegraphics[scale=1.05]{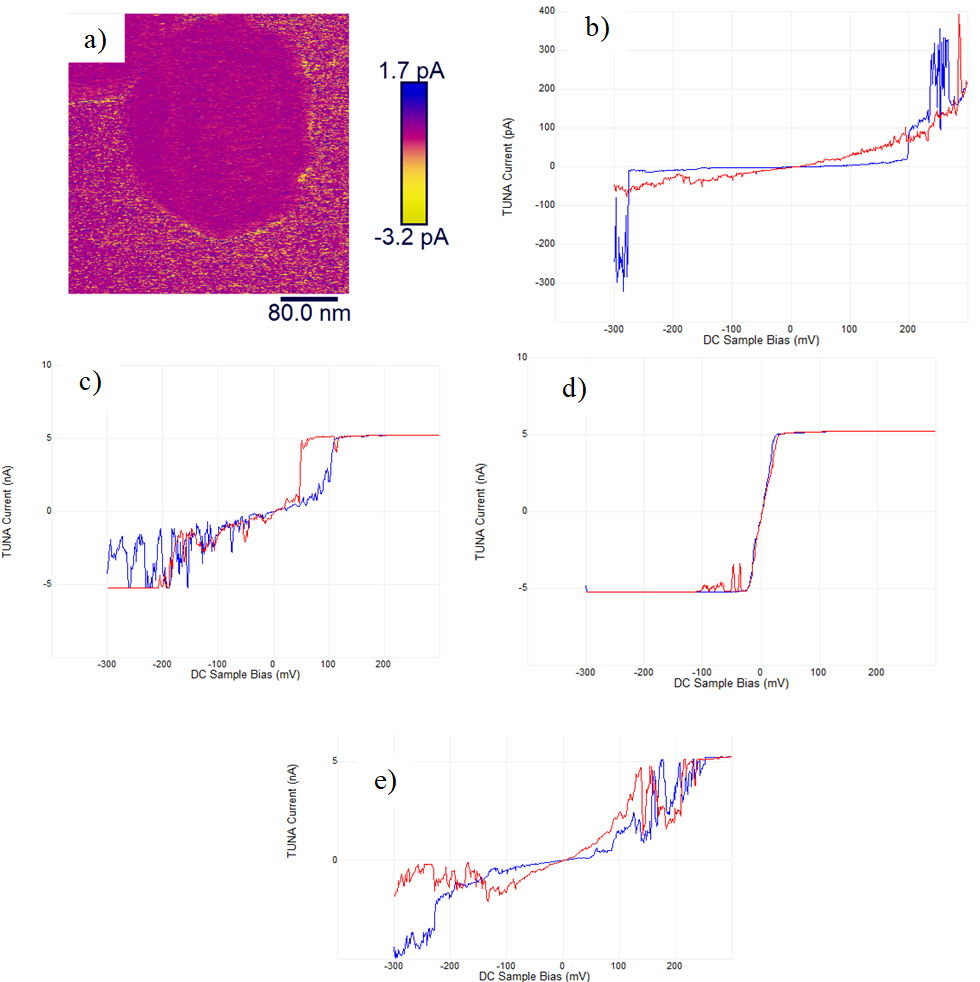}\\
\end{center}
Fig. S3. I-V spectroscopy obtained from a NW, with initially low-current characteristics. a) NW current map; b-d) I-V at increasing {contact deflection setpoints} 0.2,0.3,0.4 V; e) back to 0.25 V setpoint after the (b-d) measurements, recreating the higher resistance. These results show an increasing voltage with pressure, however it seems like contact is improving rather than a piezotronic effect. Blue and red curves are trace and retrace, correspondingly.
\end{figure}

\begin{figure}[ht]
\begin{center}
    \includegraphics[scale=1.6]{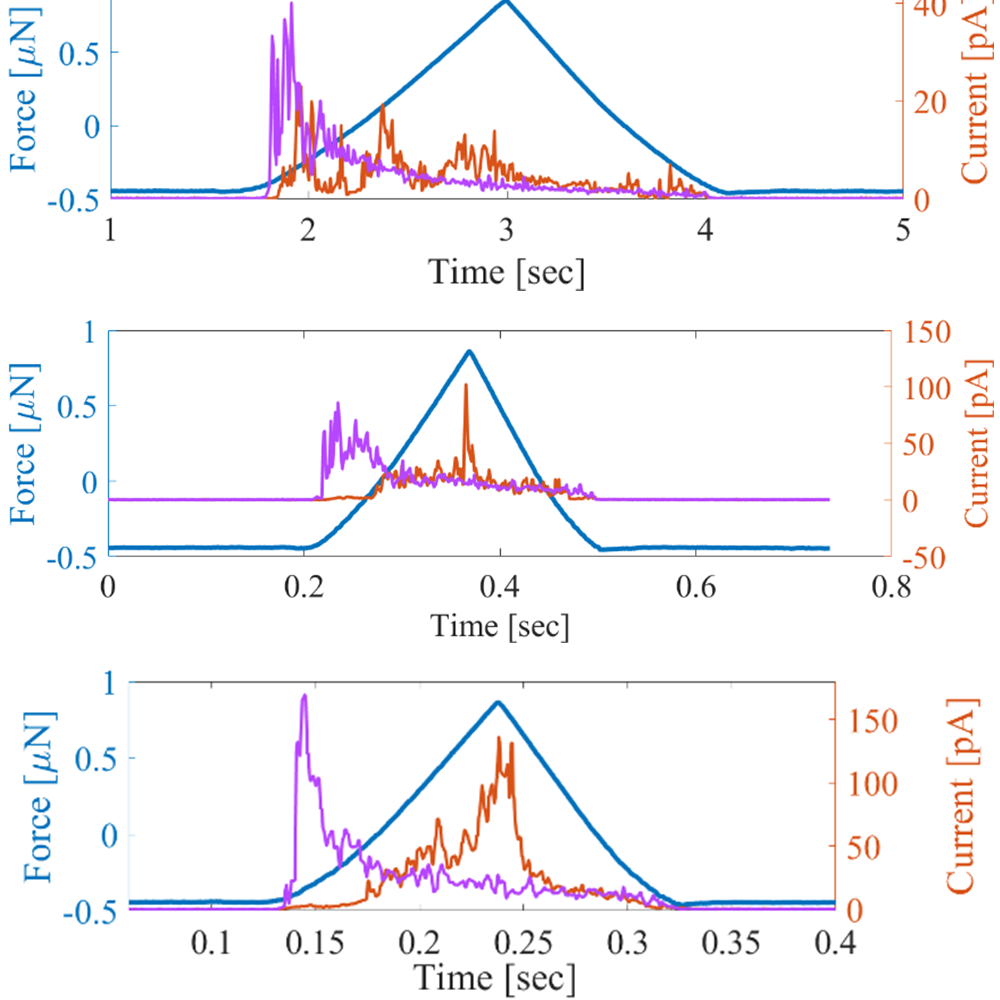}\\
\end{center}
Fig. S4. Overlaid time-resolved current measurements from main text, indicating the possible existence of a parasitic current during the measurement.
\end{figure}

\begin{figure}[ht]
\begin{center}
    \includegraphics[scale=1.6]{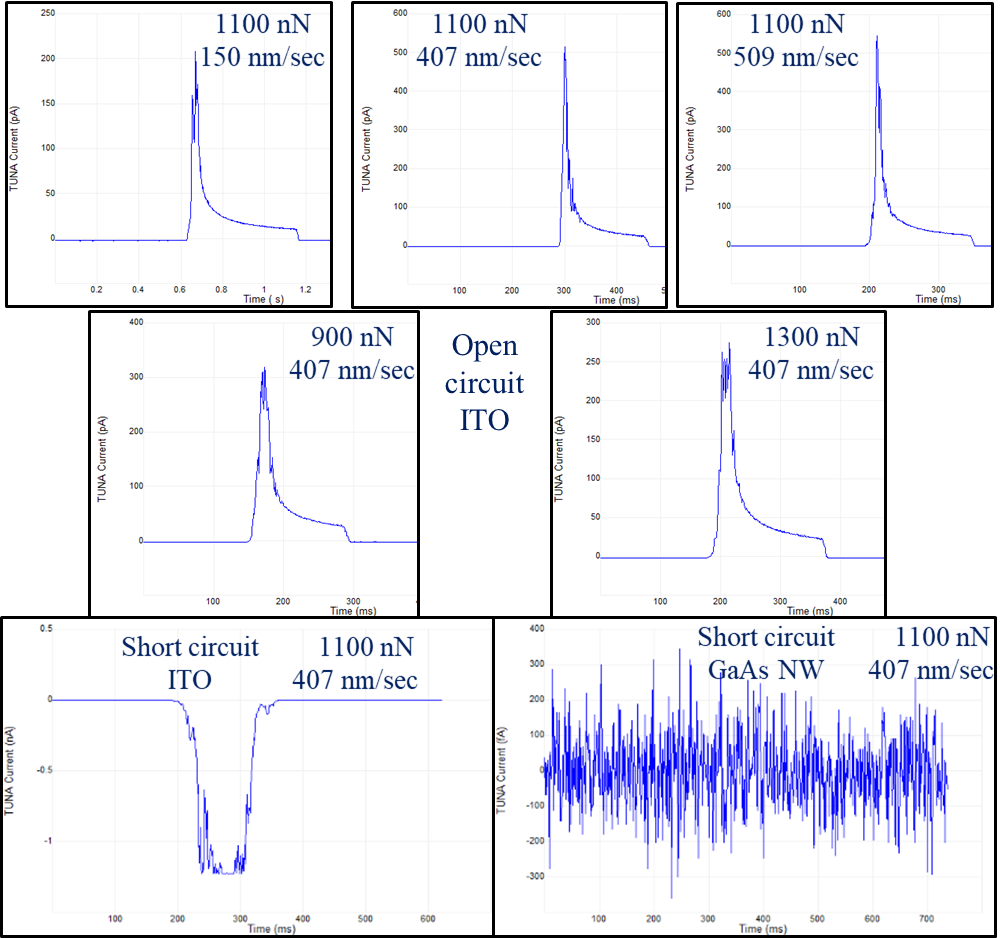}\\
\end{center}
Fig. S5. Control time-resolved current recordings: open-circuit of an ITO electrode (top five panels), with different ramp rates and max. force; short-circuit of an ITO electrode (bottom left), and a GaAs NW grown on undoped Si (bottom right). The controls show the richness of the measured current characteristics on different samples, and that there is measurement artefact – in particular in the measurement of the silicon oxide.
\end{figure}


\clearpage

\end{document}